\documentclass[a4paper,11pt]{article}
\pdfoutput=1 
\usepackage{amsmath,times,amssymb,latexsym,multicol,graphics}
\usepackage{ascmac,fancybox}
\usepackage{bbm}
\usepackage{wrapfig}%文章を画像に回り込ませる
\usepackage{framed}%文章を大枠で囲む
\usepackage{setspace} % setspaceパッケージのインクルード
\usepackage{comment}%コメント
\usepackage{authblk}%著者の情報
\usepackage{braket}
\usepackage{color}
\newcommand{\be}{\begin{equation}}
\newcommand{\ee}{\end{equation}}
\newcommand{\nn}{\nonumber}
\newcommand{\D}{\Delta}

\newcommand{\g}{\gamma}
\newcommand{\G}{\Gamma}
\newcommand{\la}{\lambda}

\newcommand{\p}{\partial}

\newcommand{\lp}{\overset{\leftarrow}{\partial}}
\newcommand{\rp}{\overset{\rightarrow}{\partial}}

\newcommand{\bpm}{\begin{pmatrix}}
\newcommand{\epm}{\end{pmatrix}}
\newcommand{\bbm}{\begin{bmatrix}}
\newcommand{\ebm}{\end{bmatrix}}

\usepackage{cite}%citeで例えば[1,2,3,4,5]を[1-5]と表示する

\usepackage{fullpage}
\begin{document} 
%%%%%%%%%%%%%%%TITLE%%%%%%%%%%%%%%%
%%%%%%%%%%%%%%%TITLE%%%%%%%%%%%%%%%
\begin{titlepage}
\begin{flushright}
{\small OU-HET 969}
 \\
\end{flushright}

\begin{center}

\vspace{1cm}

\hspace{3mm}{\LARGE \bf Fermions in Geodesic Witten Diagrams} \\[3pt] 
%\vspace{1mm}
%{\LARGE \bf }  
\vspace{1cm}

\renewcommand\thefootnote{\mbox{$\fnsymbol{footnote}$}}
Mitsuhiro {Nishida}${}^{\bar{\psi}\psi}$, Kotaro {Tamaoka}${}^{\bar{\Psi}\Psi}$

\vspace{5mm}
${}^{\bar{\psi}}${\small \sl School  of  Physics  and  Chemistry,  Gwangju  Institute of  Science  and  Technology}\\
{\small \sl Gwangju  61005,  Korea}\\
${}^{\bar{\Psi}}${\small \sl Department of Physics, Osaka University} \\ 
{\small \sl Toyonaka, Osaka 560-0043, JAPAN}\\

\vspace{5mm}
${}^{\psi}${\small{\,mnishida@gist.ac.kr}}\\
${}^{\Psi}${\small{\,k-tamaoka@het.phys.sci.osaka-u.ac.jp}
}
\end{center}

\vspace{5mm}

\noindent
\abstract
We develop the embedding formalism for odd dimensional Dirac spinors in AdS and apply it to the (geodesic) Witten diagrams including fermionic degrees of freedom. We first show that the geodesic Witten diagram (GWD) with fermion exchange is equivalent to the conformal partial waves associated with the spin one-half primary field. Then, we explicitly demonstrate the GWD decomposition of the Witten diagram including the fermion exchange with the aid of the split representation. The geodesic representation of CPW indeed gives the useful basis for computing the Witten diagrams.

\end{titlepage}
\setcounter{footnote}{0}
\renewcommand\thefootnote{\mbox{\arabic{footnote}}}
%%%%%%%%%%%%%%%TITLE%%%%%%%%%%%%%%%
\tableofcontents
\flushbottom
\section{Introduction}
The AdS/CFT correspondence\cite{Maldacena:1997re,Gubser:1998bc,Witten:1998qj} gives a plausible non-perturbative definition of the quantum gravity, via the conformal field theories (CFT). The basic building blocks in CFT are the conformal partial waves (CPW) which are the bases of correlation functions in CFT. Hence, it is natural to ask the bulk dual of CPW. Recently, the answer of this interesting question has turned out to be the geodesic Witten diagrams (GWD)\cite{Hijano:2015zsa}. See also very recent progress of the study with spinning fields\cite{Nishida:2016vds, Castro:2017hpx,Dyer:2017zef,Sleight:2017fpc,Chen:2017yia,Tamaoka:2017jce}, thermal background\cite{Kraus:2017ezw,Gobeil:2018fzy} and boundary/defect\cite{Rastelli:2017ecj,Karch:2017wgy,Sato:2017gla}. 

To show the equivalence between CPW and GWD, we can use the fact that CPW is a solution of the conformal Casimir equation. In the bulk, this equation corresponds to the equation of motion of the bulk-bulk propagator in GWD. This relation becomes manifest via the embedding formalism\cite{Dirac:1936fq,Weinberg:2010fx} by which we embed fields in AdS${}_{d+1}$/CFT${}_d$ into $\mathbb{R}^{d+1,1}$. While there is another method to solve the equation, called the shadow formalism\cite{Ferrara:1972xe,Ferrara:1972uq,Ferrara:1972ay,Ferrara:1973vz,Ferrara:1974nf,SimmonsDuffin:2012uy}, the split representation\cite{Leonhardt:2003qu,Leonhardt:2003sn,Costa:2014kfa} of the bulk-bulk propagators makes manifest the connection between them. 
It is worth noting that the embedding formalism and the split representation are also useful to compute usual Witten diagrams with or without loop effects (see, for example, \cite{Costa:2014kfa,Penedones:2010ue,Paulos:2011ie,Hikida:2017ecj,Giombi:2017hpr,Yuan:2017vgp,Giombi:2018vtc}).

While we can discuss many interesting aspects of the AdS/CFT correspondence without explicit supersymmetry, the well-known examples of the dualities are the ones between supersymmetric theories since these have been discovered from the superstring theory. Thus, it is natural to wonder if fermionic degrees of freedom might play the important role in the correspondence, for example, the stability of the vacuum at strongly coupled regime in CFT. Furthermore, superconformal partial waves are also important in the conformal bootstrap program for superconfomral field theories (see, for example, \cite{Poland:2010wg,Poland:2011ey,Beem:2013qxa}).  Therefore, studying the fermionic sector in the correspondence between CPW and GWD is quite reasonable and interesting. 

In this paper, from the above motivations, we investigate the embedding formalism for the odd dimensional AdS Dirac spinors and its application to the (geodesic) Witten diagrams. We construct 3pt and 4pt GWD with scalar and spinor fields by using the embedding formalism. Then, we check that the 4pt fermion exchange GWD with two scalar and two spinor external fields satisfies the conformal Casimir equation for CPW. To embed the AdS spinor, there is a small obstacle to impose the conventional ``transverse conditions'' such as one at the boundary. We thus need to impose a new condition that is consistent with the transverse condition at the boundary. We also derive the split representation of spinor fields and decompose the 4pt Witten diagram with fermion exchange into GWD by using the split representation. These GWD correspond to the 4pt CPW for the single and double trace operators which are constructed from scalar and spinor fields.%Note that there are a similar comment in [Weinberg], whereas we will employ a different type condition. We will see that our choice is consistent with the one at the boundary, propagators in AdS, the split representation, and so on. 

This paper is organized as follows. In section \ref{sec:emb}, we prepare the embedding formalism, the quadratic Casimir equation and the split representation for spinor fields. In section \ref{sec:gwd}, we construct the 3pt and 4pt (geodesic) Witten diagrams explicitly and show the correspondence between 4pt CPW and GWD with fermion exchange. In section \ref{sec:wd}, we decompose the 4pt Witten diagram into GWD as the CPW expansion of conformal correlation functions. We conclude with an outlook in section \ref{sec:sum}.
\section{Embedding formalism for AdS spinors}\label{sec:emb}
In this section, we develop the embedding formalism for odd dimensional AdS spinors. By using the embedding formalism, we introduce auxiliary fields, the covariant derivative, the AdS propagators and the quadratic Casimir equation for the spinor fields. We also derive the split representation of the AdS spinor bulk-bulk propagator. {Throughout this paper, we will consider Euclidean AdS and its boundary.}

\subsection{Embedding AdS${}_{d+1}$ spinors in $\mathbb{R}^{d+1,1}$}
The AdS${}_{d+1}$ space can be embedded into the $\mathbb{R}^{d+1,1}$ such that
\be
X^2=\eta_{AB}X^AX^B=-(X^0)^2+\sum^d_{i=1}(X^i)^2+(X^{d+1})^2=-X^+X^-+\sum^d_{a=1}(X^a)^2=-1.
\ee
Here $\eta_{AB}$ is the flat space metric with Lorentzian signature. In particular, we take the Poincare coordinates,
\be
X^A=(X^+,X^-,X^a)=\dfrac{1}{z}(1,z^2+x^2,x^a).
\ee
Then, the coordinates for the boundary CFT ($\mathbb{R}^d$) can be defined by its conformal boundary, 
\be
P^2=0, \;\;\;P^A\sim\la P^A (\la\in\mathbb{R}),
\ee
where
\be
P^A=(P^+,P^-,P^a)=(1,x^2,x^a).
\ee
Since we are interested in the spinor fields, we introduce the Dirac gamma matrices $\G^A$ in $\mathbb{R}^{d+1,1}$ which satisfy
\be
\left\{\G^A,\G^B\right\}=2\eta^{AB}.
\ee
In this paper, we will restrict ourself to Dirac fermions in the odd dimensional AdS space (namely, $d$ is even). It is useful to introduce an auxiliary field $S$ to contract the spinor indices such that
\be
\Psi(X,\bar{S})\equiv\bar{S}\Psi(X),
\ee
where $\bar{S}$ means Dirac conjucation of $S$. 
We would like to relate an AdS${}_{d+1}$ spinor $\psi$ with $\Psi$ in $\mathbb{R}^{d+1,1}$. Obviously, $\psi$ and $\Psi$ have different degrees of freedom in general. Thus, we need to impose some constraints for $\Psi$ so that it reduce to $\psi$ on the AdS sub-manifold. In practice, it is equivalent to say imposing the constraints for $S$. The constraint was first briefly discussed in \cite{Weinberg:2010fx}, whereas we will introduce a different condition for the AdS fermions. In particular, we impose the condition,
\be
X_A\G^A{S}_b=\G {S}_b, \label{eq:transverse}
\ee
for the auxiliary field $S$. As we spell out below, our basis of the fermions \eqref{eq:transverse} makes the connection to the original AdS ones manifest.
Here $\G$ is the chiral gamma matrix in $\mathbb{R}^{d+1,1}$. Notice that we can {\it not} take the conventional ``transverse condition'', $X_A\G^A{S}_b=0$ for the AdS spinors. Such a naive condition gives rise to ${S}_b=0$ due to the non-vanishing determinant of $X_A\G^A$. In our convention, summarized in Appendix \ref{app:spinor}, the solution of \eqref{eq:transverse} is
\be
S_b=\dfrac{1}{\sqrt{z}}\bbm (z\g^0+\g^ax_a)\chi \\ \chi\ebm, \label{eq:indexsb}
\ee
where $\g^a\, (a=1,2, \cdots d)$ is the $d$-dimensional gamma matrices, and $\g^0$ is the chiral gamma matrix in the same dimension. Here $\chi$ is an auxiliary field for the original AdS. 
Taking the limit $z\rightarrow0$, $S_b$ goes to $z^{-\frac{1}{2}}S_\p$, where
%Similarly, we define an auxiliary field $S_\partial$ for the boundary spinor as
\begin{align}
S_\p=\bbm (\g^ax_a)s \\ s\ebm.
\end{align}
Here $s$ is an auxiliary field in the boundary CFT. This $S_\p$ is conventional one in the embedding formalism for CFT fermions\cite{Iliesiu:2015qra,Iliesiu:2015akf,Isono:2017grm}.
\subsection{Equation of motion and its solutions}
We shall define the covariant derivative $\nabla_A$ for the embedding fermions. 
It should be the straightforward extension for the tensor fields. That is,
\be
\nabla_A=G_{AB}\p^B+\Sigma_{AB}X^B,
\ee
where $G_{AB}\equiv \eta_{AB}+X_AX_B$ is the induced metric, and $\Sigma_{AB}\equiv \frac{1}{4}[\G^A,\G^B]$ is the generator of rotation for the fermions. Then, the Dirac equation in the embedding space is \footnote{With the index-free notation, these derivatives $\Gamma^A\nabla_A$ should be replaced by $\bar{S}_b\Gamma^A\nabla_A\p_{\bar{S}_b}$.}
\begin{align}
\left[\G^A\nabla_A-m\right]\Psi(X)&=0, \label{eq:Dirac}\\ 
\bar{\Psi}(X)\left[\overset{\leftarrow}{\nabla}_A\G^A+m\right]&=\left[(G_{AB}(\p^B\bar{\Psi}(X))-\bar{\Psi}(X)\Sigma_{AB}X^B)\G^A+m\bar{\Psi}(X)\right]=0.\label{eq:bDirac}
\end{align}
Hereafter we will use the notation $\overset{\leftarrow}{\,}$ that represents the differential action from the right side.
\subsubsection{Propagators}
One can explicitly check that there are non-normalizable solutions of \eqref{eq:Dirac}, namely the bulk-boundary propagators for the spinors\cite{Henningson:1998cd,Kawano:1999au}\footnote{Our definiton of $\D$ is the same as one of \cite{Henningson:1998cd}, however, it is different from one of \cite{Kawano:1999au}.}. In the embedding space, these are given by
\begin{align}
G^{\D,\frac{1}{2}}_{b\p}(X,\bar{S}_b;P,S_\p)&=\mathcal{C}_{\D,\frac{1}{2}}\dfrac{\braket{\bar{S}_b\Pi_-S_\p}}{(-2X\cdot P)^{\D+\frac{1}{2}}},\label{bbp1}\\
\bar{G}^{\D,\frac{1}{2}}_{b\p}(X,S_b;P,\bar{S}_\p)&=\mathcal{C}_{\D,\frac{1}{2}}\dfrac{\braket{\bar{S}_\p\Pi_-S_b}}{(-2X\cdot P)^{\D+\frac{1}{2}}},\label{bbp2}
\end{align}
where
\be
\mathcal{C}_{\D,\frac{1}{2}}=\dfrac{1}{\pi^{d/2}}\dfrac{\G\left(\D+\frac{1}{2}\right)}{\G\left(\D+\frac{1-d}{2}\right)},
\ee
is the normalization constant fixed by the behavior at the boundary. 
Here $\Pi_\pm$ is the chiral projection, $\Pi_\pm\equiv\frac{1\pm\G}{2}$. We also used the relation $m=\D-\frac{d}{2}$. Note that the chiral projection is necessarily for the propagators to be the solutions. In the AdS/CFT correspondence, on-shell Dirac fermions in the AdS space become Weyl fermions in the boundary theory\cite{Henningson:1998cd}. If we flip the sign of the mass $m$ to $-m$ in the equation of motion (\ref{eq:Dirac}), the chiral projection $\Pi_-$ in (\ref{bbp1}) also becomes $\Pi_+$. For the notation and connection to the usual AdS space, please see appendix \ref{app:spinor}. {The bulk-boundary propagator for the scalar fields is given by
\begin{align}
G^{\D,0}_{b\p}(X;P)&=\mathcal{C}_{\D,0}\frac{1}{(-2X\cdot P)^{\D}},\\
\mathcal{C}_{\D,0}&=\dfrac{1}{2\pi^{d/2}}\dfrac{\G\left(\D\right)}{\G\left(\D+1-\frac{d}{2}\right)}.
\end{align}}

We also have the bulk-bulk propagator\cite{Kawano:1999au} given by
\be
G^{\D,\frac{1}{2}}_{bb}(X,\bar{S}_b;Y,T_b)=\braket{\bar{S}_b\,\Pi_+T_b}\left(\dfrac{d}{du}G^{\D_+,0}_{bb}(u)\right)+\braket{\bar{S}_b\,\Pi_-T_b}\left(\dfrac{d}{du}G^{\D_-,0}_{bb}(u)\right),\label{bbp3}
\ee
where $G^{\D,0}_{bb}(u)$ is the bulk-bulk propagator for a scalar field, and
\begin{align}
\dfrac{d}{du}G^{\D,0}_{bb}(u)&=-\dfrac{1}{\pi^h}\dfrac{\G(\D+1)}{\G(\D-h+1)}(2u)^{-1}F_{2,\D}(u),\label{bbp4}\\
F_{2,\D}(u)&=(2u)^{-\D}{}_2F_1(\D+1,\D-h+1/2,2\D-2h+1,-2u^{-1})\nn\\
&=\frac{u}{2^\D(u+1)^{\D+1}}{}_2F_1\left(\frac{\D+1}{2},\frac{\D}{2}+1,\D-h+1,\frac{1}{(u+1)^2}\right).
\end{align}
Here we used $\D_\pm=\D\pm1/2$, $h=d/2$ and the chordal distance $u=-1-X\cdot Y$.

\subsubsection{Connection to the conformal Casimir equation}
As usual, we also have the Klein-Gordon type equation of motion,
\be
\left[(\Gamma^A\nabla_A)^2-m^2\right]\Psi(X)=0.
\ee
By using the above definitions, one can show the well-known relation
\be
(\Gamma^A\nabla_A)^2=\eta^{AB}\nabla_A\nabla_B-\dfrac{1}{4}R,\label{DO}
\ee
where $R=-d(d+1)$ is Ricci scalar for AdS${}_{d+1}$. Thus, the Klein-Gordon equation is now equivalent to
\be
\left[\nabla^2-m^2_{\textrm{eff.}}\right]\Psi(X)=0, 
\ee
where $m^2_{\textrm{eff.}}=m^2-\frac{R}{4}$ is an effective mass. 
From the above equation, we obtain a useful relation for the AdS Dirac fermion,
\be
-\dfrac{1}{2}L^{AB}L_{AB}\Psi(X)=\left[(\G^A\nabla_A)^2+\dfrac{1}{8}R\right]\Psi(X)=C_{\D,\frac{1}{2}}\Psi(X),\label{CE}
\ee
where $L_{AB}=X_A\p^X_B-X_B\p^X_A+\Sigma_{AB}$ is the $SO(d+1,1)$ generator. Here $C_{\D,\frac{1}{2}}$ is the same as the quadratic conformal Casimir for $d$-dimensional spinor representation, $C_{\D,\frac{1}{2}}=\D(\D-d)+\frac{d}{8}(d-1)$, {(not the coefficient of the bulk-boundary propagators $\mathcal{C}_{\D,\frac{1}{2}}$)}. 
In particular, the bulk-bulk propagator \eqref{bbp3} satisfies the above quadratic Casimir equation \eqref{CE} when $X\neq Y$.
This fact will be important to show that GWD satisfies the conformal Casimir equation. 

\subsection{Split representation}
Let us define the harmonic function for the AdS fermions,
\be
\Omega_{\nu,\frac{1}{2}}(X,Y)\equiv\dfrac{i}{2\pi}\int_{\p \textrm{AdS}_{d+1}} [dP]\Bigg[G^{h+i\nu,\frac{1}{2}}_{b\p}(X,\bar{S}_b;P,S_\p)\Bigg](\lp_{S_\p}P\rp_{\bar{S}_\p})\Bigg[\left.\bar{G}^{h-i\nu,\frac{1}{2}}_{b\p}(Y,S_b;P,\bar{S}_\p)\right|_{\Pi_+\leftrightarrow\Pi_-}\Bigg],\label{eq:omega}
\ee
where $\lp_{S}P\rp_{\bar{S}}$ is the projector and necessary for the conformal integral with respect to $P$\cite{SimmonsDuffin:2012uy}. In what follows, we will often use a short hand notation for the Dirac matrices, say $P\equiv P^A\G_A$. \eqref{eq:omega} is a solution of the Dirac equation,
\be
[(\Gamma\cdot\nabla_X)-i\nu]\Omega_{\nu,\frac{1}{2}}(X,Y)=\Omega_{\nu,\frac{1}{2}}(X,Y)[(\overset{\leftarrow}{\nabla}_{Y}\cdot\Gamma)-i\nu]=0.
\ee
Note that $\Omega_{\nu,\frac{1}{2}}$ is {\it not} a solution of the other Dirac equation $\bar{\Psi}[(\overset{\leftarrow}{\nabla}_X\cdot\Gamma)+i\nu]=0$.
One can confirm this feature by showing that $\Omega_{\nu,\frac{1}{2}}$ is a linear combination of the bulk-bulk propagators with different scaling dimensions (see appendix \ref{app:SR}),
\be
\Omega_{\nu,\frac{1}{2}}(X,Y)=\dfrac{i}{2\pi}\left\{G^{h+i\nu,\frac{1}{2}}_{bb}(X,Y)-\Bigg(\left.G^{h-i\nu,\frac{1}{2}}_{bb}(X,Y)\right|_{\Pi_+\leftrightarrow\Pi_-}\Bigg)\right\}.\label{sr1} 
\ee
Then, the split representation of the fermion bulk-bulk propagator is,
\be
G^{\D,\frac{1}{2}}_{bb}(X,Y)=\int^{\infty}_{-\infty} \dfrac{d\nu}{\nu+i(\D-h)}\,\Omega_{\nu,\frac{1}{2}}(X,Y).\label{eq:split}
\ee
The derivation of (\ref{eq:split}) is as follows. From the explicit form of the propagator, $G^{h\pm i\nu,\frac{1}{2}}_{bb}$ converges at $\textrm{Im}(\nu)\to \mp\infty$. Then, we can show\footnote{We implicitly assume $\D>h=d/2$.}
\begin{align}
&\dfrac{i}{2\pi}\int^{\infty}_{-\infty} \dfrac{d\nu}{\nu+i(\D-h)}\,G^{h+i\nu,\frac{1}{2}}_{bb}(X,Y)=G^{\D,\frac{1}{2}}_{bb}(X,Y),\notag\\
&\dfrac{i}{2\pi}\int^{\infty}_{-\infty} \dfrac{d\nu}{\nu+i(\D-h)}\,G^{h-i\nu,\frac{1}{2}}_{bb}(X,Y)|_{\Pi_+\leftrightarrow\Pi_-}=0,\label{sr2} 
\end{align}
from the residue theorem. These lead (\ref{eq:split}).

\section{Conformal partial waves from geodesic Witten diagrams}\label{sec:gwd}
In this section, we show the equivalence between 4pt CPW and GWD including fermionic degrees of freedom. In section \ref{subsec:3ptgwd}, we start analysis of 3pt tree-level (geodesic) Witten diagrams with two spinor fields and a scalar field. We highlight the ratio of these amplitude, which will be useful in the next section. In section \ref{subsec:4ptgwd}, we demonstrate the aforementioned correspondence especially between the 4pt CPW and GWD including an internal spinor field and external scalar and spinor fields.

	\subsection{Warm up: Fermion-Fermion-Scalar amplitude}\label{subsec:3ptgwd}
	As warm up exercise, let us compute the tree-level 3pt Witten diagram associated with the Yukawa-like interaction in the embedding space, 
	\be
	S_{int.}=\int_{\textrm{AdS$_{d+1}$}}dX\,\bar{\Psi}_1(X)\Psi_2(X)\Phi_3(X). \label{eq:yukawa}
	\ee
	The amplitude of the diagram can be written as
	\be
	\mathcal{A}_3=\int_{\textrm{AdS}_{d+1}} dX\,\bar{G}^{\D_1,\frac{1}{2}}_{b\p}(X,S_b;P_1,\bar{S}_{1\p})(\lp_{S_b}\rp_{\bar{S}_b})G^{\D_2,\frac{1}{2}}_{b\p}(X,\bar{S}_b;P_2,S_{2\p})G^{\D_3,0}_{b\p}(X;P_3).
	\ee
	It reduces to
	\be
	\mathcal{A}_3=\braket{\bar{S}_{1\p}\Pi_-S_{2\p}}\mathcal{C}_{\D_1,\frac{1}{2}}\mathcal{C}_{\D_2,\frac{1}{2}}\mathcal{C}_{\D_3,0}\int_{\textrm{AdS}_{d+1}} dX\,\dfrac{1}{(-2P_1\cdot X)^{\delta_1}(-2P_2\cdot X)^{\delta_2}(-2P_3\cdot X)^{\delta_3}}. 
	\ee
Here we introduced $\delta_i\equiv\D_i+s_i$, where $s_i$ is magnitude of the spin of $i$-th operator. (Namely, $\delta_1=\D_1+\frac{1}{2}, \delta_2=\D_2+\frac{1}{2}$ and $\delta_3=\D_3$ in this example.) Notice that the integrand is the same as the scalar 3pt amplitude of the Witten diagram with scaling dimension $\delta_{1},\delta_2$ and $\delta_3$. By using the standard Schwinger-Feynman integral method, one can obtain
\begin{align}
\mathcal{A}_3&=C_{\bar{\psi}_1\psi_2\mathcal{O}_3}\dfrac{\braket{\bar{S}_{1\p}\Pi_- S_{2\p}}}{(-2P_1\cdot P_2)^{\frac{1}{2}(\delta_1+\delta_2-\delta_3)}(-2P_2\cdot P_3)^{\frac{1}{2}(\delta_2+\delta_3-\delta_1)}(-2P_3\cdot P_1)^{\frac{1}{2}(\delta_3+\delta_1-\delta_2)}},\\
C_{\bar{\psi}_1\psi_2\mathcal{O}_3}&=\pi^h\mathcal{C}_{\D_1,\frac{1}{2}}\mathcal{C}_{\D_2,\frac{1}{2}}\mathcal{C}_{\D_3,0}\G\left(\dfrac{1}{2}\left(-d+\sum^3_{i=1}\delta_i\right)\right)\nn\\
&\hspace{2cm}\times\dfrac{\G\left(\frac{1}{2}(\delta_1+\delta_2-\delta_3)\right)\G\left(\frac{1}{2}(\delta_2+\delta_3-\delta_1)\right)\G\left(\frac{1}{2}(\delta_3+\delta_1-\delta_2)\right)}{2\G(\delta_1)\G(\delta_2)\G(\delta_3)}.
\end{align}
Then, let us consider the corresponding 3pt {\it geodesic} Witten diagram, 
	\be
	\mathcal{W}_3(\g_{12})=\int_{\g_{12}} d\la\,\bar{G}^{\D_1,\frac{1}{2}}_{b\p}(X,S_b;P_1,\bar{S}_{1\p})(\lp_{S_b}\rp_{\bar{S}_b})G^{\D_2,\frac{1}{2}}_{b\p}(X,\bar{S}_b;P_2,S_{2\p})G^{\D_3,0}_{b\p}(X;P_3).\label{eq:3ptgwd1}
	\ee
	Here the integration domain $\g_{ij}$ represents the geodesic anchored on the boundary points $P_i$ and $P_j$. The bulk coordinate X on the geodesic $\g_{ij}$ is given by
	\be
	X_A(\la)=\dfrac{e^{-\la}P_{1A}+e^{\la}P_{2A}}{\sqrt{-2P_1\cdot P_2}}. 
	\ee
	By using the integral representation of the beta function, we readily obtain
	\begin{align}
	\mathcal{W}_3(\g_{12})&=C^{(\g_{12})}_{\bar{\psi}_1\psi_2\mathcal{O}_3}\dfrac{\braket{\bar{S}_{1\p}\Pi_- S_{2\p}}}{(-2P_1\cdot P_2)^{\frac{1}{2}(\delta_1+\delta_2-\delta_3)}(-2P_2\cdot P_3)^{\frac{1}{2}(\delta_2+\delta_3-\delta_1)}(-2P_3\cdot P_1)^{\frac{1}{2}(\delta_3+\delta_1-\delta_2)}},\label{eq:ffs121}\\
	C^{(\g_{12})}_{\bar{\psi}_1\psi_2\mathcal{O}_3}&=\dfrac{1}{2}\mathcal{C}_{\D_1,\frac{1}{2}}\mathcal{C}_{\D_2,\frac{1}{2}}\mathcal{C}_{\D_3,0}\,B\left(\frac{\delta_3-\delta_1+\delta_2}{2},\frac{\delta_3+\delta_1-\delta_2}{2}\right). \label{eq:ffs122}
	\end{align}
	It is worth noting that, the ``ratio'' of the usual diagram $\mathcal{A}_3$ to the geodesic one $\mathcal{W}_3(\g_{12})$ is
	\be
	\dfrac{\mathcal{A}_3}{\mathcal{W}_3(\g_{12})}=\pi^h\G\left(\dfrac{1}{2}\left(-d+\sum^3_{i=1}\delta_i\right)\right)\dfrac{\G\left(\frac{1}{2}(\delta_1+\delta_2-\delta_3)\right)}{\G(\delta_1)\G(\delta_2)}
	\ee
	In the same way, one can evaluate $\mathcal{W}_3(\g_{23})$ and $\mathcal{W}_3(\g_{31})$. As expected, we get the same 3pt function as \eqref{eq:ffs121}, while the over all coefficients are different from $\mathcal{W}_3(\g_{12})$:
	\begin{align}
	\dfrac{\mathcal{A}_3}{\mathcal{W}_3(\g_{23})}&=\pi^h\G\left(\dfrac{1}{2}\left(-d+\sum^3_{i=1}\delta_i\right)\right)\dfrac{\G\left(\frac{1}{2}(\delta_2+\delta_3-\delta_1)\right)}{\G(\delta_2)\G(\delta_3)},\\
	\dfrac{\mathcal{A}_3}{\mathcal{W}_3(\g_{31})}&=\pi^h\G\left(\dfrac{1}{2}\left(-d+\sum^3_{i=1}\delta_i\right)\right)\dfrac{\G\left(\frac{1}{2}(\delta_3+\delta_1-\delta_2)\right)}{\G(\delta_3)\G(\delta_1)}.
	\end{align}
One can also obtain the above 3pt functions with opposite chirality by replacing $\Pi_\mp\rightarrow\Pi_\pm$ for each intermediate step. For $\Psi(P_1)$ with the chirality $\Pi_-$, generic fermion-fermion-scalar 3pt correlators have the following two different tensor structures:
\be
\braket{\Psi(P_1,\bar{S}_1)\bar{\Psi}(P_2,S_2)\Phi(P_3)}=\dfrac{C_1\braket{\bar{S}_{1\p}\Pi_- S_{2\p}}+C_2\braket{\bar{S}_{1\p}\Pi_-P_3S_{2\p}}\sqrt{\frac{(-2P_1\cdot P_2)}{(-2P_1\cdot P_3)(-2P_2\cdot P_3)}}}{(-2P_1\cdot P_2)^{\frac{1}{2}(\delta_1+\delta_2-\delta_3)}(-2P_2\cdot P_3)^{\frac{1}{2}(\delta_2+\delta_3-\delta_1)}(-2P_3\cdot P_1)^{\frac{1}{2}(\delta_3+\delta_1-\delta_2)}}\label{eq:gen3pt_ffs},
\ee
where $C_{1,2}$ are OPE coefficients for each spinor bi-linear. In the same manner, we have two classes of 3pt interactions in AdS for fermion-fermion-scalar vertices. To get the second term of \eqref{eq:gen3pt_ffs} from the AdS integral, for example, one can consider a derivative interaction:
\be
S_{int.}=\int_{\textrm{AdS$_{d+1}$}}dX\,\bar{\Psi}_1(X)\G^A\Psi_2(X)\nabla_A\Phi_3(X).\label{eq:derivint}
\ee
One can also take a more exotic interaction,
\be
S_{int.}=\int_{\textrm{AdS$_{d+1}$}}dX\,\bar{\Psi}_1(X)\G^A\Psi_2(X)X_A\Phi_3(X),\label{eq:int2}
\ee
which will be formally appeared in the GWD decomposition of the Witten diagram in the next section. In these cases, we need to take the bulk-boundary propagators with relatively opposite chirality projection because of $\G^A$ in (\ref{eq:derivint}) and (\ref{eq:int2}). {In appendix \ref{subsec:3ptderiv}, we leave intermediate steps from the derivative interaction \eqref{eq:derivint} to the second spinor bi-linear in \eqref{eq:gen3pt_ffs}. We will compute 3pt GWD with the exotic one \eqref{eq:int2} in section \ref{sec:wd}.
}

\subsection{Conformal partial waves from geodesic diagrams with fermion exchange}\label{subsec:4ptgwd}
Next, we would like to show the equivalence between CPW and GWD both of which possess external/internal fermions. However, GWD with 4 external fermions and an internal scalar is almost trivially equivalent to CPW. One may understand this triviality from the previous 3pt calculations --- (\ref{eq:3ptgwd1}) can be expressed as a scalar 3pt GWD times the spinor bi-linear $\braket{\bar{S}_{1\p}\Pi_-S_{2\p}}$. Similarly, the above 4pt GWD can be written as a scalar 4pt GWD times spinor bi-linears such as $\braket{\bar{S}_{1\p}\Pi_-S_{2\p}}$ and $\braket{\bar{S}_{3\p}\Pi_-S_{4\p}}$. Therefore, the problem almost reduces to the correspondence between the scalar GWD and CPW. 

Thus, here we display only the detailed proof for the fermion exchange. 
We consider a fermion exchange GWD with two external spinors and two external scalars in the embedding space. The amplitude of this GWD is given by
\be
\mathcal{W}_4(\D,\D_i)=\int^\infty_{-\infty} d\la F_\D[P_1,P_2,Y(\la),\bar{S}_{1\p},T_b](\lp_{T_b}(1+Y(\la))\rp_{\bar{T}_b})G^{\D_4,\frac{1}{2}}_{b\p}(Y(\la),\bar{T}_{b};P_4, S_{4\p})G^{\D_3,0}_{b\p}(Y(\la);P_3),\label{eq:gwd}
\ee
where
\begin{align}
&F_\D[P_1,P_2,Y,\bar{S}_{1\p},T_b]\nn\\
&\equiv \int^\infty_{-\infty} d\la\, \bar{G}^{\D_1,\frac{1}{2}}_{b\p}(X(\la),S_{b};P_1, \bar{S}_{1\p})G^{\D_2,0}_{b\p}(X(\la);P_2)(\lp_{S_b}(1+X(\la))\rp_{\bar{S}_b})G^{\D,\frac{1}{2}}_{bb}(X(\la),\bar{S}_b;Y,T_b).\label{F1} 
\end{align}
The reader may wonder why we consider a strange interaction $(\lp_{S_b}(1+X)\rp_{\bar{S}_b})$ rather than $(\lp_{S_b}\rp_{\bar{S}_b})$. 
As discussed below, one can also take the latter. The different choice gives rise to the different combination of spinor bi-linears in the corresponding CPW.
For the meanwhile, however, we will take the former (strange one) since we can show that this choice of interaction reproduces the Yukawa-like interaction for the original AdS${}_{d+1}$ space,
\be
\int_{\textrm{AdS}_{d+1}} \sqrt{-g}d^{d+1}x  \;\bar{\psi}_{\D_i}\psi_{\D_j}\phi_{\D_k}.\label{YI} 
\ee
In other words, our choice of the interaction will be useful for the bulk computation in the next section\footnote{Unfortunately, the relation of 3pt interactions between the embedding space and the original space is non-trivial in general. This troublesomeness simply comes from the higher dimensional embedding of spinors \eqref{eq:indexsb}.}. {For more detail, please see appendix \ref{subsec:4yukawa}.}

The above function $F[P_1,P_2,Y,\bar{S}_{1\p},T_b]$ is invariant under the simultaneous rotation by $L_1+L_2+\overset{\leftarrow}{L_Y}$. Using this fact twice and \eqref{CE}, we get
	\begin{align}
	-\dfrac{1}{2}(L_1+L_2)^2 F_\D[P_1,P_2,Y,\bar{S}_{1\p},T_b]&=F_\D[P_1,P_2,Y,\bar{S}_{1\p},T_b]\left(-\dfrac{1}{2}(\overset{\leftarrow}{L_Y})^2\right)\nn\\
	&=C_{\D,\frac{1}{2}} F_\D[P_1,P_2,Y,\bar{S}_{1\p},T_b]. 
	\end{align}
Therefore, we have shown that $\mathcal{W}_4$ satisfies the conformal Casimir equation:
\be
-\dfrac{1}{2}(L_1+L_2)^2\mathcal{W}_4(\D,\D_i)=C_{\D,\frac{1}{2}}\mathcal{W}_4(\D,\D_i).\label{eq:ccegwd}
\ee
Moreover, the above GWD has two independent solutions of the conformal Casimir equation since $L_{AB}=Y_A\p^Y_B-Y_B\p^Y_A+\Sigma_{AB}$ commutes with $\Pi_\pm$. Let us decompose the above GWD $\mathcal{W}_4$ into two pieces,
\begin{align}
\mathcal{W}_4(\D,\D_i)&=\mathcal{W}^{+}_4(\D,\D_i)+\mathcal{W}^{-}_4(\D,\D_i),\label{GWD}\\
\mathcal{W}^{\pm}_4(\D,\D_i)&\equiv\int^\infty_{-\infty} d\la F^{\pm}_\D[P_1,P_2,Y(\la),\bar{S}_{1\p},T_b](\lp_{T_b}(1+Y(\la))\rp_{\bar{T}_b})G^{\D_4,\frac{1}{2}}_{b\p}(Y(\la),\bar{T}_{b};P_4, S_{4\p})G^{\D_3,0}_{b\p}(Y(\la);P_3),\label{eq:gwdpm}
\end{align}
where we defined
\begin{align}
F^{\pm}_\D[P_1,P_2,Y,\bar{S}_{1\p},T_b]&\nn\\
&\hspace{-2.9cm}\equiv \int^\infty_{-\infty} d\la\, \bar{G}^{\D_1,\frac{1}{2}}_{b\p}(X(\la),S_{b};P_1, \bar{S}_{1\p})G^{\D_2,0}_{b\p}(X(\la);P_2)(\lp_{S_b}(1+X(\la))\rp_{\bar{S}_b})G^{\D,\frac{1}{2},\pm}_{bb}(X(\la),\bar{S}_b;Y,T_b), \\
G^{\D,\frac{1}{2},\pm}_{bb}(X,\bar{S}_b;Y,T_b)&=\braket{\bar{S}_b\,\Pi_\pm T_b}\left(\dfrac{d}{du}G^{\D_\pm,0}_{bb}(u)\right).\label{eq:pmgbb}
\end{align}
Both of $\mathcal{W}^{\pm}_4(\D,\D_i)$ indeed satisfy the appropriate boundary conditions; hence, $\mathcal{W}^{\pm}_4(\D,\D_i)$ are individually CPW with different spinor bi-linears $\braket{S_{1\p}\Pi_-P_2P_3S_{4\p}}$ and $\braket{S_{1\p}\Pi_-S_{4\p}}$, respectively\footnote{{Here each chiral projection in \eqref{eq:pmgbb} selects one of the spinor bi-linears. For example, $W^+_4(\D,\D_i)$ initially includes 4 spinor bi-linears $\braket{S_{1\partial}\Pi_{-}1\Pi_{+}1\Pi_{-}S_{4\partial}}$, $\braket{S_{1\partial}\Pi_{-} X \Pi_{+}1\Pi_{-}S_{4\partial}}$, $\braket{S_{1\partial}\Pi_{-}1\Pi_{+}Y\Pi_{-}S_{4\partial}}$, and $\braket{S_{1\partial}\Pi_{-} X \Pi_{+}Y\Pi_{-}S_{4\partial}}$. 
The first three bi-linears become zero, and only $\braket{S_{1\partial}\Pi_{-} X \Pi_{+}Y\Pi_{-}S_{4\partial}}=\braket{S_{1\partial}\Pi_{-} XYS_{4\partial}}
$ survives. Then, this leads to $\braket{S_{1\partial}\Pi_{-} P_2P_3S_{4\partial}}$.}}. To extract one of these bi-linears from GWD, one needs other ``3pt interactions'' for its amplitude, whereas the above choice is obviously convenient for the GWD decomposition of the Witten diagram with (\ref{YI}). For example, if we consider $(\lp_{S_b}\rp_{\bar{S}_b})$  instead of $(\lp_{S_b}(1+X)\rp_{\bar{S}_b})$ in (\ref{F1}), (\ref{eq:gwd}) reduces to $\mathcal{W}^{-}_4(\D,\D_i)$. 

Finally, we comment on the extension to the external higher spin fields. 
We can construct these CPW from the seed CPW with differential operators invented in \cite{Costa:2011dw}. Similarly, we can construct the corresponding GWD from the seed GWD as studied in \cite{Nishida:2016vds,Castro:2017hpx,Dyer:2017zef,Sleight:2017fpc,Chen:2017yia}. We can follow the similar story even with fermions. In our example, $\mathcal{W}^{\pm}_4$ play the role of the seed GWDs. However, the relation between the differential operators in the boundary and 3pt interactions in the bulk would be involved. We will come back this point in section \ref{sec:sum}. 

\section{GWD decomposition of Witten diagram}\label{sec:wd}
Finally, we demonstrate the GWD decomposition of an exchange Witten diagram with fermion exchange as a concrete example. {Based on the method in \cite{Chen:2017yia}, we can extract the conformal dimension in the CPW expansion systematically from the ratio between the usual diagram $\mathcal{A}_3$ and the geodesic one $\mathcal{W}_3$.}

We shall consider the following amplitude,
\begin{align}
\mathcal{A}_4&=\int dX\int dY\; \bar{G}^{\D_1,\frac{1}{2}}_{b\p}(X,S_{b};P_1, \bar{S}_{1\p})G^{\D_2,0}_{b\p}(X;P_2)\nn\\
&\times(\lp_{S_b}(1+X)\rp_{\bar{S}_b})G^{\D_0,\frac{1}{2}}_{bb}(X,\bar{S}_b;Y,T_b)(\lp_{T_b}(1+Y)\rp_{\bar{T}_b})G^{\D_4,\frac{1}{2}}_{b\p}(Y,\bar{T}_{b};P_4, S_{4\p})G^{\D_3,0}_{b\p}(Y;P_3).\label{eq:wd}
\end{align}
The difference between \eqref{eq:wd} and the previous amplitude \eqref{eq:gwd} is just the integration domain, the geodesics or the entire bulk. We would like to decompose \eqref{eq:wd} into the summation of \eqref{eq:gwd} and to read off the OPE data. 
To this end, it is useful to employ the shadow formalism. By using \eqref{eq:split}, we obtain the relation between 3pt and 4pt GWD as
\begin{align}
\int_{\p \textrm{AdS}_{d+1}} [dP] \mathcal{W}_3(P_1,P_2,P)(\lp_{S_\p}P\rp_{\bar{S}_\p}) \mathcal{W}^\prime_3(P,P_3,P_4)&=\sum_{c=\pm}\big[\mathcal{W}_4^c(\D,\D_i)-\mathcal{W}^{\textrm{shadow},c}_4(\D,\D_i)\big],\label{eq:shadow1}
\end{align}
where\footnote{{Based on the cubic vertices in (\ref{eq:wd}), we can also define $\mathcal{W}_3$ by using $\lp_{S_b}(1+X)\rp_{\bar{S}_b}$ instead of $\lp_{S_b}\rp_{\bar{S}_b}$. Such definition is reduced to (\ref{eq:splitgwd1}) owing to the chiral projection operators $\Pi_\pm$ in the fermion propagators. The same is true for $\mathcal{W}_3^\prime$.}}
\begin{align}
&\mathcal{W}_3(P_1,P_2,P)=\int_{\g_{12}} d\la\,\bar{G}^{\D_1,\frac{1}{2}}_{b\p}(X(\la),S_b;P_1,\bar{S}_{1\p})(\lp_{S_b}\rp_{\bar{S}_b})G^{\D,\frac{1}{2}}_{b\p}(X(\la),\bar{S}_b;P,S_{\p})G^{\D_2,0}_{b\p}(X(\la);P_2),\label{eq:splitgwd1}\\
&\mathcal{W}^\prime_3(P,P_3,P_4)=\int_{\g_{34}} d\la\,\left.\bar{G}^{d-\D,\frac{1}{2}}_{b\p}(X(\la),S_b;P,\bar{S}_{\p})\right|_{\Pi_+\leftrightarrow\Pi_-}\hspace{-1cm}(\lp_{S_b}X(\la)\rp_{\bar{S}_b})G^{\D_4,\frac{1}{2}}_{b\p}(X(\la),\bar{S}_b;P_4,S_{4\p})G^{\D_3,0}_{b\p}(X(\la);P_3),\label{eq:splitgwd2}\end{align}
and
\begin{align}
 \mathcal{W}^{\textrm{shadow},\pm}_4(\D,\D_i)&\equiv\int^\infty_{-\infty} d\la \left[F^{\pm}_{d-\D}[P_1,P_2,Y(\la),\bar{S}_{1\p},T_b]\Big|_{\Pi_\pm\rightarrow\Pi_{\mp}}\right]\nn\\
 &\times(\lp_{T_b}(1+Y(\la))\rp_{\bar{T}_b})G^{\D_4,\frac{1}{2}}_{b\p}(Y(\la),\bar{T}_{b};P_4, S_{4\p})G^{\D_3,0}_{b\p}(Y(\la);P_3). \label{Ws}
\end{align}
We also define $\mathcal{A}_3$ and $\mathcal{A}^\prime_3$ by replacing the geodesic integrals of \eqref{eq:splitgwd1} and \eqref{eq:splitgwd2} with the entire bulk integrals. The former ones, $\mathcal{A}_3$ and $\mathcal{W}_3$, were computed in the previous section. Here we note the final result of $\mathcal{A}^\prime_3$ and $\mathcal{A}^\prime_3/\mathcal{W}^\prime_3$ for references;
\begin{align}
\mathcal{A}^\prime_3&=\dfrac{\mathcal{C}_{d-\D,\frac{1}{2}}\mathcal{C}_{\D_3,0}\mathcal{C}_{\D_4,\frac{1}{2}}}{2\tau}\braket{\bar{S}_{\p}\Pi_+\dfrac{\p}{\p P}S_4}\int dX\;\dfrac{1}{(-2X\cdot P)^{\tau}(-2X\cdot P_3)^{\delta_3}(-2X\cdot P_4)^{\delta_4}}\nn\\
&=\dfrac{\tau+\delta_3-\delta_4}{2\tau}\dfrac{\pi^h}{2}\G\left(\dfrac{1}{2}\left(-d+\tau+\delta_3+\delta_4\right)\right)\dfrac{\G(\frac{1}{2}(\tau+\delta_3-\delta_4))\G(\frac{1}{2}(\delta_3+\delta_4-\tau))\G(\frac{1}{2}(\delta_4+\tau-\delta_3))}{\G(\tau)\G(\delta_3)\G(\delta_4)}\nn\\
&\times\mathcal{C}_{d-\D,\frac{1}{2}}\mathcal{C}_{\D_3,0}\mathcal{C}_{\D_4,\frac{1}{2}}\dfrac{\braket{\bar{S}_\p\Pi_+P_3S_4}}{(-2P\cdot P_3)^{\frac{1}{2}(\tau+\delta_3-\delta_4)+1}(-2P_3\cdot P_4)^{\frac{1}{2}(\delta_3+\delta_4-\tau)}(-2P\cdot P_4)^{\frac{1}{2}(\tau+\delta_4-\delta_3)}},\label{eq:A3prime}
\end{align}
\begin{align}
\dfrac{\mathcal{A}^\prime_3}{\mathcal{W}^\prime_3}=\pi^h\G\left(\dfrac{1}{2}\left(-d+\tau+\delta_3+\delta_4\right)\right)\dfrac{\G\left(\dfrac{1}{2}\left(\delta_3+\delta_4-\tau\right)\right)}{\G(\delta_3)\G(\delta_4)},\label{eq:A3/W3prime}
\end{align}
where we defined $\tau\equiv d-\D-\frac{1}{2}=h-i\nu-\frac{1}{2}$.
Then, we can rewrite (\ref{eq:wd}) by using 4pt GWD,
\begin{align}
\mathcal{A}_4&=\frac{i}{2\pi}\int \dfrac{d\nu}{\nu+i(\D_0-h)}\int_{\p \textrm{AdS}_{d+1}} [dP] \mathcal{A}_3(P_1,P_2,P)(\lp_{S_\p}P\rp_{\bar{S}_\p}) \mathcal{A}^\prime_3(P,P_3,P_4)\nn\\
&=\frac{i}{2\pi}\int \dfrac{d\nu}{\nu+i(\D_0-h)}(\mathcal{A}_3/\mathcal{W}_3)(\mathcal{A}^\prime_3/\mathcal{W}^\prime_3)\int_{\p \textrm{AdS}_{d+1}} [dP] \mathcal{W}_3(P_1,P_2,P)(\lp_{S_\p}P\rp_{\bar{S}_\p}) \mathcal{W}^\prime_3(P,P_3,P_4)\notag\\
&={\frac{i}{2\pi}}\int \dfrac{d\nu}{\nu+i(\D_0-h)}(\mathcal{A}_3/\mathcal{W}_3)(\mathcal{A}^\prime_3/\mathcal{W}^\prime_3)\left(\mathcal{W}_4^++\mathcal{W}_4^--\mathcal{W}^{\textrm{shadow},+}_4-\mathcal{W}^{\textrm{shadow},-}_4\right),\label{GWDe}
\end{align}
where $\mathcal{W}_4^\pm$ includes $G^{\D,\frac{1}{2},\pm}_{bb}=G^{h+ i\nu,\frac{1}{2},\pm}_{bb}$, and $\mathcal{W}^{\textrm{shadow},\pm}_4$ does $G^{d-\D,\frac{1}{2},\pm}_{bb}=G^{h- i\nu,\frac{1}{2},\pm}_{bb}$.

We apply a complex contour integral for the integration with respect to $\nu$ in (\ref{GWDe}).
Since $G^{h\pm i\nu,\frac{1}{2},\pm}_{bb}$ converges at $\textrm{Im}(\nu)\to \mp\infty$, we consider the lower half complex $\nu$-plane for $\mathcal{W}_4^\pm$ and the upper half complex $\nu$-plane for $\mathcal{W}^{\textrm{shadow},\pm}_4$. In the lower half plane, $\dfrac{(\mathcal{A}_3/\mathcal{W}_3)(\mathcal{A}^\prime_3/\mathcal{W}^\prime_3)}{\nu+i(\D_0-h)}$ has the poles at
\begin{align}
&h+i\nu=\D_0,\label{st}\\
&h+i\nu=\Delta_1+\Delta_2+2m,\;\;\; h+i\nu=\Delta_3+\Delta_4+2m, \;\;\;(m=0, 1, 2, \cdots).\label{dt1}
\end{align}
In the upper half plane, the poles of $\dfrac{(\mathcal{A}_3/\mathcal{W}_3)(\mathcal{A}^\prime_3/\mathcal{W}^\prime_3)}{\nu+i(\D_0-h)}$ are
\begin{align}
&h-i\nu=\Delta_1+\Delta_2+2m+1,\;\;\; h-i\nu=\Delta_3+\Delta_4+2m+1, \;\;\;(m=0, 1, 2, \cdots).\label{dt2}
\end{align}
In the point of the view of CFT, (\ref{st}) corresponds to the single trace spinor operator. (\ref{dt1}) and (\ref{dt2}) correspond to the double trace operators which are constructed from the scalar and spinor fields. {Their schematic forms are $\phi_{\D_i}(\partial_a\partial^a)^m\psi_{\D_j}+\dots$, where $\phi_{\D_i}$ and $\psi_{\D_j}$  are the scalar and spinor fields in CFT.}  We note that the difference between $2m$ and $2m+1$ in (\ref{dt1}) and (\ref{dt2}) is related to $\Pi_+\leftrightarrow\Pi_-$ exchange in (\ref{Ws}). For example, $G^{\D_1+\D_2+2m,\frac{1}{2},+}_{bb}(X,\bar{S}_b;Y,T_b)$ and $G^{\D_1+\D_2+2m+1,\frac{1}{2},-}_{bb}(X,\bar{S}_b;Y,T_b)|_{\Pi_+\leftrightarrow\Pi_-}$ are the same propagators. One can obtain the coefficients of $\mathcal{W}_4$ by applying the residue theorem explicitly.

Summarizing the above, the 4pt Witten diagram (\ref{eq:wd}) can be expressed by GWD as (\ref{GWDe}). After the integration with respect to $\nu$,  we can obtain the GWD expansion of  (\ref{eq:wd}). These GWD correspond to CPW for the single trace spinor operator with the conformal dimension (\ref{st}) and the double trace operators with the conformal dimension (\ref{dt1}). Generalization to the GWD decomposition of other Witten diagrams is straightforward, but we leave it for future work.

\section{Outlook}\label{sec:sum}
In this paper, we have developed the embedding formalism for AdS fermions in section \ref{sec:emb} and applied it to the proof of the correspondence between GWD and CPW with fermions in section\ref{sec:gwd}. We have also decomposed the fermion exchange Witten diagram into the infinite sum of GWD (equivalently, CPW) in section \ref{sec:wd}. The key ingredient for the computation was the split representation for AdS fermions, which was given by \eqref{eq:split}. We list several future directions as concluding remarks. 

\paragraph{Generalization to odd $d$}
In this paper, we focused on the embedding formalism with the even dimensional Minkowski spacetime.  It is also important to organize the embedding formalism in the odd dimensional Minkowski spacetime for the odd dimensional CPW and the even dimensional GWD. In odd dimensional field theories, we cannot introduce chirality or Weyl fermions. Thus, the spinor structure of the independent even dimensional GWD is probably different from the odd dimensional one. Interesting research topics in the odd dimensional CFT with fermion are, for example, 3d (supersymmetric) $U(N)$ model\cite{Gross:1974jv,Moshe:2003xn,Hikida:2016cla,Hikida:2017ecj} and 1d cSYK\cite{Gross:2017vhb,Gross:2017aos}.

\paragraph{Make geodesic Witten diagram super}
One of the motivation of our work was to extend the arguments to the superconformal field theories (SCFT). We can decompose correlation functions of SCFT into the superconformal partial waves (SCPW) corresponding to the exchange of superconformal primaries and their descendants. The superconformal partial waves are the solutions of the {\it super} conformal Casimir equation. The most transparent approach should be again the {\it super} embedding formalism which has been used for constructing SCPW via the shadow formalism\cite{Fitzpatrick:2014oza}. Then, it is quite reasonable to develop the counterpart in AdS fields. It would help to construct the super version of GWD and show their equivalence to SCPW.

\paragraph{Higher spin fields and interactions}
One may be curious about extending the argument in the present paper to higher (half-integer) spin fields, for example, the gravitino. With the previous developments for the bosonic higher spin fields\cite{Hijano:2015zsa,Nishida:2016vds,Castro:2017hpx,Dyer:2017zef,Sleight:2017fpc,Chen:2017yia,Costa:2014kfa,Tamaoka:2017jce}, one can straightforwardly extend our calculation to the (geodesic) Witten diagrams with the most generic representations. However, the calculation does not become so simple since the spinor bi-linears in the embedding AdS do not satisfy the transverse condition. In addition, the relation between 3pt interactions in the original space and ones in the embedding space becomes involved due to the higher dimensional embedding of fermions; but of course, one can formally introduce 3pt interactions in the embedding space and use it for GWD. This is because the interaction for GWD is not unique and probably no physical meaning. Improving these unsatisfactory points might give more useful tools to compute the Witten diagrams including higher spin fermionic degrees of freedom.

\section*{Acknowledgements }
We would like to thank Yasuaki Hikida, Hideki Kyono and Takahiro Nishinaka for useful comments and discussions.  KT would also like to thank the KIAS-YITP joint workshop 2017 ``Strings, Gravity and Cosmology'' where a part of this work was presented. The work of MN was supported by Basic Science Research Program through the National Research Foundation of Korea (NRF) funded by the Ministry of Science, ICT \& Future Planning (NRF- 2017R1A2B4004810) and GIST Research Institute (GRI) grant funded by the GIST in 2018.

\appendix
\section{Convention for spinors}\label{app:spinor}
In this appendix, we summarize our convention of spinors in $d+2$ dimensional Lorentzian spacetime (the embedding space for AdS${}_{d+1}$/CFT${}_d$). For more detail, we refer to \cite{Iliesiu:2015qra,Iliesiu:2015akf,Isono:2017grm}. We basically follows one in \cite{Isono:2017grm}.

The gamma matrices $\G^A$ are now given by
\be
\G^0=\bbm 0 & 1\\-1&0\ebm, \G^a=\bbm \g^a & 0\\0&-\g^a\ebm, \G^{d+1}=\bbm 0 & 1\\1&0\ebm,
\ee
where $\g^a$s satisfy $\{\g^a, \g^b\}=\delta^{ab}$ ($a,b=1,2,\cdots,d$). 
The chiral gamma matrix for $d+2=2n+2$ dimension is defined by
\be
\G\equiv\dfrac{1}{i^{n+3}}\G^0\G^1\cdots\G^{d+1}\equiv\bbm-\g^0&0\\0&\g^0\ebm,
\ee
where we defined one for $2n$-dimension (with Euclidian signature) as $\g^0\equiv -i^{-(n+1)}\g^1\g^2\cdots\g^d$. For the Dirac conjugation, we use $-\G^0$ rather than $\G^0$ so does \cite{Isono:2017grm}. Namely, 
\be
\bar{\Psi}\equiv\Psi^\dagger(-\Gamma^0).
\ee

Next, we briefly note the convention for the boundary fermions. The primary fields with spin $\frac{1}{2}$ should satisfy
\be
\Psi_\p(\la P)=\la^{-(\D+\frac{1}{2})}\Psi_\p(P), 
\ee
so that it can reproduce the conformal algebra for spin $\frac{1}{2}$ fields.
We also introduce the index-free notation for the boundary spinors, 
\be
\Psi(P,\bar{S}_\p)\equiv\bar{S}_\p\Psi(P). 
\ee
On the projective null cone $P^2=0$, $S_\p$ satisfies the conventional ``transverse condition'', 
\be
\G^AP_AS_\p=0.
\ee
We can consistently impose this condition since $(\G^AP_A)^2=0$, as opposed to the bulk fermions. We also use the notation for the spinor bi-linears for the auxiliary fields, say $\braket{\bar{S}_{b1}\cdots S_{\p2}}$. Here the suffixes $b$ ($\p$) represent the fields on the bulk (boundary). 

We leave some explicit formulae of the spinor bi-linears in the original space,
\begin{align}
\braket{\bar{S}_b\Pi_\pm S_{\p}}&=-\dfrac{1}{\sqrt{z}}\chi^\dagger[z\g^0+\g^a(x_b-x_\p)_a]\mathcal{P}_\pm s,\\
\braket{\bar{S}_\p\Pi_\pm S_b}&=\dfrac{1}{\sqrt{z}}s^\dagger\mathcal{P}_\mp[z\g^0+\g^a(x_b-x_\p)_a]\chi,\\
\braket{\bar{S}_{bi}\Pi_\pm S_{bj}}&=\dfrac{1}{\sqrt{z_iz_j}}\chi^\dagger_i[\mathcal{P}_{\mp}\g^\mu (z_{j})_{\mu}-\g^\mu (z_{i})_{\mu}\mathcal{P}_{\pm}]\chi_j,
\end{align}
where $\chi$ ($s$) are the bulk (boundary) auxiliary fields in the original space, and $\mathcal{P}_{\mp}\equiv\frac{1\mp\g^0}{2}$. {With these formulae, the AdS propagators (\ref{bbp1}), (\ref{bbp2}) and (\ref{bbp3}) can be written as
\begin{subequations}\label{subeq:propagators}
\begin{align}
G^{\D,\frac{1}{2}}_{b\p}(z,x_b,\chi;x_\p,s)&=-\dfrac{\mathcal{C}_{\D,\frac{1}{2}}}{\sqrt{z}}\chi^\dagger[\g^0z+\g^a(x_b-x_\p)_a]\mathcal{P}_-s\left(\dfrac{z}{z^2+(x_b-x_\p)^2}\right)^{\D+\frac{1}{2}},\\
\bar{G}^{\D,\frac{1}{2}}_{b\p}(z,x_b,\chi;x_\p,s)&=\dfrac{\mathcal{C}_{\D,\frac{1}{2}}}{\sqrt{z}}s^\dagger\mathcal{P}_+[\g^0z+\g^a(x_b-x_\p)_a]\chi\left(\dfrac{z}{z^2+(x_b-x_\p)^2}\right)^{\D+\frac{1}{2}},\\
G^{\D,\frac{1}{2}}_{bb}(z_1,x_{1b},\chi_1;z_2,x_{2b},\chi_2)&=-\dfrac{1}{\sqrt{z_1z_2}}\Big[\chi^\dagger_1(z_1^\mu\g_\mu\mathcal{P}_--\mathcal{P}_+z_2^\mu\g_\mu)\chi_2\dfrac{d}{du}G^{\D_-,0}_{bb}(u)\notag\\
&\;\;\;\;\;\;\;\;\;\;\;\;\;\;\;+\chi^\dagger_1(z_1^\mu\g_\mu\mathcal{P}_+-\mathcal{P}_-z_2^\mu\g_\mu)\chi_2\dfrac{d}{du}G^{\D_+,0}_{bb}(u)\Big].
\end{align}
\end{subequations}
After dropping the auxiliary fields $s$ and $\chi$, these are consistent with the expressions in \cite{Kawano:1999au}.}

We also leave several formulae about the covariant derivative for help to check (\ref{DO}) and (\ref{CE}):
\begin{align}
\G^A\nabla_A\G^B\nabla_B&=\G^A[\nabla_A,\G^B]\nabla_B+\eta^{AB}\nabla_A\nabla_B+2\Sigma^{AB}\nabla_A\nabla_B,\\
\G^A[\nabla_A,\G^B]\nabla_B&=X^A\G_A(\Gamma\cdot\nabla)=+\Sigma^{AB}L^X_{AB}-\dfrac{d+1}{2},\\
[G_{AC}\p^C_X,G_{BD}\p^D_X]&=-L^X_{AB},\\
2\Sigma^{AB}[\Sigma_{AC}X^C,G_{BD}\p^D_X]&=-\dfrac{R}{2}+\dfrac{(d+1)}{2},\\
\Sigma^{AB}[\Sigma_{AC},\Sigma_{BD}]X^CX^D&=\dfrac{R}{4},\\
\eta^{AB}\nabla_A\nabla_B&=[G_{AB}\p^A_X\p^B_X+(d+1)(X\cdot\p_X)]-L^X_{AB}\Sigma^{AB}+\eta^{AB}\Sigma_{AC}\Sigma_{BD}X^CX^D,\\
\eta^{AB}\Sigma_{AC}\Sigma_{BD}&=-\dfrac{1}{4}\left[d\,\G_C\G_D+\eta_{CD}\right],\\
\Sigma^{AB}\Sigma_{AB}&=\dfrac{1}{4}R-\dfrac{d+1}{2},
\end{align}
where $R=-d(d+1)$ and $L_{AB}^X=X_A\p^X_B-X_B\p^X_A$.

\section{Computational details}
In this appendix, we supply some of the details which were omitted in the main text for the sake of presentation. 
\subsection{Derivation of (\ref{sr1})}\label{app:SR}
We complement the intermediate calculation of (\ref{sr1}). In \cite{Hikida:2017ecj}, there is the same analysis by Fourier transformation without the embedding formalism. Let us start to rewrite the harmonic function (\ref{eq:omega}),
\be
\Omega_{\nu,\frac{1}{2}}(X,Y)=\frac{i}{2\pi}\mathcal{C}_{\D,\frac{1}{2}}\mathcal{C}_{d-\D,\frac{1}{2}}\int_{\p \textrm{AdS}_{d+1}} [dP] \dfrac{\braket{\bar{S}_b\Pi_-S_\p}}{(-2X\cdot P)^{\D+\frac{1}{2}}}(\lp_{S_\p}P\rp_{\bar{S}_\p})\dfrac{\braket{\bar{S}_\p \Pi_+T_b}}{(-2P\cdot Y)^{(d-\D)+\frac{1}{2}}},\label{App1}
\ee
where $\D=h+i\nu$. 
By using the following formulae:
\begin{align}
\int_{\p AdS_{d+1}}[dP]\;\dfrac{P_A}{[-2P\cdot Y]^{d+1}}&=\dfrac{\pi^{h}\G(h+1)}{\G(2h+1)}\dfrac{Y_A}{(-Y^2)^{h+1}}, \label{eq:conf1}\\
\frac{1}{A^xB^y}&=\frac{\G(x+y)}{\G(x)\G(y)}\int^\infty_0\frac{dt}{t}t^y\frac{1}{[A+tB]^{x+y}},\label{eq:Feynman2}
\end{align}
we obtain
\begin{align}
\Omega_{\D,\frac{1}{2}}(X,Y)&=\frac{i}{2\pi}\mathcal{C}_{\D,\frac{1}{2}}\mathcal{C}_{d-\D,\frac{1}{2}}\dfrac{\pi^h\G(h+1)}{\G(\D+\frac{1}{2})\G(d-\D+\frac{1}{2})}\int^\infty_0\dfrac{dt}{t}t^{d-\D+\frac{1}{2}}\dfrac{\braket{\bar{S}_b\,\Pi_+T_b}-t\braket{\bar{S}_b\,\Pi_-T_b}}{[-(X+tY)^2]^{h+1}},\label{App4}
\end{align}
where we used (\ref{eq:transverse}). The 1st term in (\ref{App4}) is
\begin{align}
&\;\;\;\;\;\mathcal{C}_{\D,\frac{1}{2}}\mathcal{C}_{d-\D,\frac{1}{2}}\dfrac{\pi^h\G(h+1)}{\G(\D+\frac{1}{2})\G(d-\D+\frac{1}{2})}\braket{\bar{S}_b\,\Pi_+T_b}\int^\infty_0\dfrac{dt}{t}t^{d-\D+\frac{1}{2}}\dfrac{1}{[-(X+tY)^2]^{h+1}}\nn\\
&=\braket{\bar{S}_b\,\Pi_+T_b}\Bigg[\left(\dfrac{d}{du}G^{\D_+,0}_{bb}(u)\right)-\left(\dfrac{d}{du}G^{\widetilde{\D}_+,0}_{bb}(u)\right)\Bigg],
\end{align}
where we used
\begin{align}
\int^\infty_0\dfrac{dt}{t}\dfrac{t^{-c}}{\left(\frac{(1+t)^2}{t}+2u\right)^b}&=\dfrac{\G(b+c)\G(-c)}{\G(b)(2u)^{b+c}}{}_2F_1\left(b+c,\frac{1}{2}+c,1+2c,-\frac{2}{u}\right)\nn\\
&+\dfrac{\G(b-c)\G(c)}{\G(b)(2u)^{b-c}}{}_2F_1\left(b-c,\frac{1}{2}-c,1+2c,-\frac{2}{u}\right),\label{eq:int2F1}
\end{align}
and we defined $\D_{\pm}\equiv \D\pm\frac{1}{2}$ and its shadow $\widetilde{\D}_{\pm}\equiv d-\D_{\pm}=d-\D\mp\frac{1}{2}$. The 2nd term in (\ref{App4}) also becomes 
\begin{align}
&\mathcal{C}_{\D,\frac{1}{2}}\mathcal{C}_{d-\D,\frac{1}{2}}\dfrac{\pi^h\G(h+1)}{\G(\D+\frac{1}{2})\G(d-\D+\frac{1}{2})}\braket{\bar{S}_b\,\Pi_-T_b}\int^\infty_0\dfrac{dt}{t}t^{d-\D+\frac{1}{2}}\dfrac{(-t)}{[-(X+tY)^2]^{h+1}}\nn\\
&=\braket{\bar{S}_b\,\Pi_-T_b}\Bigg[\left(\dfrac{d}{du}G^{\D_-,0}_{bb}(u)\right)-\left(\dfrac{d}{du}G^{\widetilde{\D}_-,0}_{bb}(u)\right)\Bigg].
\end{align}
Therefore, we obtain (\ref{sr1}). From the similar computation, one can also derive (\ref{eq:shadow1}).

\subsection{3pt amplitude with a derivative interaction}\label{subsec:3ptderiv}
Here we compute 3pt GWD with derivative interaction \eqref{eq:derivint},
\be
S_{int.}(\g_{12})=\int_{\textrm{AdS$_{d+1}$}}dX\,\bar{\Psi}_1(X)\G^A\Psi_2(X)\nabla_A\Phi_3(X).\nn
\ee
In this case, 3pt GWD is given by
\begin{align}
\mathcal{W}_{3,deriv}(\g_{12})&=\int_{\g_{12}} d\la\,\left.\bar{G}^{\D_1,\frac{1}{2}}_{b\p}(X(\la),S_b;P_1,\bar{S}_{1\p})(\lp_{S_b}\Gamma^A\rp_{\bar{S}_b})\Big[G^{\D_2,\frac{1}{2}}_{b\p}(X(\la),\bar{S}_b;P_2,S_{2\p})\right|_{\Pi_+\leftrightarrow\Pi_-}\Big]\nn\\
&\hspace{9cm}\times\nabla_AG^{\D_3,0}_{b\p}(X(\la);P_3)\\
&\propto 2\D_3\int_{\g_{12}} d\la\,\dfrac{\braket{\bar{S}_{1\p}\Pi_-P_3S_{2\p}}}{(-2P_1\cdot X(\la))^{\D_1+\frac{1}{2}}(-2P_2\cdot X(\la))^{\D_2+\frac{1}{2}}(-2P_3\cdot X(\la))^{\D_3+1}}\\
&\propto\dfrac{\braket{\bar{S}_{1\p}\Pi_-P_3S_{2\p}}\sqrt{\frac{(-2P_1\cdot P_2)}{(-2P_1\cdot P_3)(-2P_2\cdot P_3)}}}{(-2P_1\cdot P_2)^{\frac{1}{2}(\delta_1+\delta_2-\delta_3)}(-2P_2\cdot P_3)^{\frac{1}{2}(\delta_2+\delta_3-\delta_1)}(-2P_3\cdot P_1)^{\frac{1}{2}(\delta_3+\delta_1-\delta_2)}}.
\end{align}
Hence, we have obtained the second spinor bi-linear in \eqref{eq:gen3pt_ffs}.
\subsection{Yukawa-like interaction: embedding space v.s. physical space}\label{subsec:4yukawa}
Finally, we note some intermediate steps to relate the strange interaction in the embedding AdS space with Yukawa-like interaction in the {\it physical} one. 
In the embedding space, the exchange diagram \eqref{eq:wd} reduces to 
\begin{align}
\mathcal{A}_4&=\mathcal{C}_{\D_1,\frac{1}{2}}\mathcal{C}_{\D_2,0}\mathcal{C}_{\D_3,\frac{1}{2}}\mathcal{C}_{\D_4,0}\int dX\int dY\dfrac{\braket{\bar{S}_{1\p}\Pi_{-}S_{3\p}}\dfrac{d}{du}G^{\D_-}_{bb}(u)+\braket{\bar{S}_{1\p}\Pi_-X\Pi_{+}Y\Pi_{-}S_{3\p}}\dfrac{d}{du}G^{\D_+}_{bb}(u)}{(-2P_1\cdot X)^{\D_1+\frac{1}{2}}(-2P_2\cdot X)^{\D_2}(-2P_3\cdot Y)^{\D_3+\frac{1}{2}}(-2P_4\cdot Y)^{\D_4}}\nn\\
&=\mathcal{C}_{\D_1,\frac{1}{2}}\mathcal{C}_{\D_2,0}\mathcal{C}_{\D_3,\frac{1}{2}}\mathcal{C}_{\D_4,0}\int dX\int dY\dfrac{\braket{\bar{S}_{1\p}\Pi_{-}S_{3\p}}\dfrac{d}{du}G^{\D_-}_{bb}(u)+\braket{\bar{S}_{1\p}\Pi_-XYS_{3\p}}\dfrac{d}{du}G^{\D_+}_{bb}(u)}{(-2P_1\cdot X)^{\D_1+\frac{1}{2}}(-2P_2\cdot X)^{\D_2}(-2P_3\cdot Y)^{\D_3+\frac{1}{2}}(-2P_4\cdot Y)^{\D_4}}\label{eq:a4reduced}
\end{align}
One can see that this amplitude is equivalent to the one in the physical space with the propagators \eqref{subeq:propagators} and Yukawa-like interaction \eqref{YI} (for each vertecies),
\begin{align}
\mathcal{A}^{phys.}_4&=\int \dfrac{dz_1d^{d}x_{1b}}{z^{d+1}_1}\int \dfrac{dz_2d^{d}x_{2b}}{z^{d+1}_2}\;\Big[\bar{G}^{\D_1,\frac{1}{2}}_{b\p}(z_1,x_{1b};x_{1\p})G^{\D_2,0}_{b\p}(z_1,x_{1b};x_{2\p})\nn\\
&\hspace{2cm}\times G^{\D,\frac{1}{2}}_{bb}(z_1,x_{1b};z_2,x_{2b})G^{\D_3,\frac{1}{2}}_{b\p}(z_1,x_{2b};x_{3\p})G^{\D_4,0}_{b\p}(z_1,x_{2b};x_{4\p})\Big],
\end{align}
where the above propagators are given by \eqref{subeq:propagators} (peeled off the auxiliary spinors). For the explicit check, it is useful to note that
\begin{align}
&\braket{\bar{S}_{1\p}\Pi_-X_1X_2S_{2\p}}\nn\\
&=\dfrac{1}{z_1z_2}s_1^\dagger[\mathcal{P}_+\{(z^2_1(x_{2b}-x_{1\p})_i\gamma^i-z^2_2(x_{1b}-x_{1\p})^a\gamma_a+(x_{1b}-x_{1\p})_i(x_{1b}-x_{2b})_j(x_{2b}-x_{2\p})_k\gamma^i\gamma^j\gamma^k)\}]s_2.
\end{align}
Notice that the second term in \eqref{eq:a4reduced} gives rise to the spinor bi-linear $\braket{\bar{S}_{1\p}\Pi_{-}P_2P_4S_{3\p}}$. This can be easily seen in the GWD,
\be
\mathcal{W}_4\propto\int_{\gamma_{12}} d\la\int_{\gamma_{34}} d\la^\prime\dfrac{\braket{\bar{S}_{1\p}\Pi_{-}S_{3\p}}\dfrac{d}{du}G^{\D_-}_{bb}(u)+\braket{\bar{S}_{1\p}\Pi_-X(\la)Y(\la^\prime)S_{3\p}}\dfrac{d}{du}G^{\D_+}_{bb}(u)}{(-2P_1\cdot X(\la))^{\D_1+\frac{1}{2}}(-2P_2\cdot X(\la))^{\D_2}(-2P_3\cdot Y(\la^\prime))^{\D_3+\frac{1}{2}}(-2P_4\cdot Y(\la^\prime))^{\D_4}}.
\ee
For $\mathcal{W}_4$, it is obvious that $\braket{\bar{S}_{1\p}\Pi_-XYS_{3\p}}$ becomes proportional to $\braket{\bar{S}_{1\p}\Pi_-P_2P_4S_{3\p}}$ on the geodesics $\gamma_{12}$ and $\gamma_{34}$. 
%%%

\end{document}